\documentstyle[12pt]{article}

\begin{document}

\begin{center}
{\bf The Large Numbers Hypothesis and Quantum Mechanics}
\end{center}

\begin{center}
Saulo Carneiro
\end{center}

\begin{center}
{\small \it Instituto de F\'{\i}sica, Universidade Federal da Bahia\\
40210-340, Salvador, BA, Brasil}
\end{center}

\begin{abstract}

In this paper, the suggested similarity between micro and
macro-cosmos is extended to quantum behavior, postulating that
quantum mechanics, like general relativity and classical
electrodynamics, is invariant under discrete scale transformations. This
hypothesis leads to a large scale quantization of angular momenta. Using
the scale factor
$\Lambda \sim 10^{38}$, the corresponding quantum of action,
obtained by scaling the Planck constant, is close to the Kerr limit
for the spin of the universe -- when this is considered as a huge
rotating black-hole -- and to the spin of Godel's universe, solution
of Einstein equations of gravitation. Besides, we suggest the existence
of another, intermediate, scale invariance, with scale factor
$\lambda \sim 10^{19}$. With this factor we obtain, from Fermi's
scale, the values for the gravitational radius and for the collapse
proper-time
of a typical black-hole, besides the Kerr limit value for
its spin. It is shown that the mass-spin relations implied by the two
referred scale transformations are in accordance with
Muradian's Regge-like relations for galaxy clusters and stars.
Impressive results are derived when we use a $\lambda$-scaled quantum 
approach to calculate the mean radii of planetary orbits in solar system.
Finally, a possible explanation for the
observed quantization of galactic redshifts is suggested, based on the
large scale quantization conjecture.

\end{abstract}

\section{Introduction}

${ }$

One of the oldest curious features of particles physics and cosmology
is the possibility of obtaining cosmological large numbers, as
mass ($M$), radius ($R$) and age ($T$) of our universe, scaling up the
typical values of mass ($m$), size ($r$) and life-time ($t$) appearing
in particles physics, by a scale factor $\Lambda \sim 10^{38-41}$.

This fact has led to important ideas and developments, as
Dirac's hypothesis that cosmological parameters vary with the age of
universe$^{[1]}$ and strong gravity$^{[2-5]}$, that tries to derive the
hadron properties from a scaling down of gravitational theory, treating
particles as black-hole type solutions.

The strong gravity approach can be based on the scale invariance of
general relativity, also present in classical electrodynamics: the
gravitational and electromagnetic equations are invariant under a
scale transformation of time intervals and distances, provided we
scale too the correspondent coupling factors. With this philosophy, we
can think the universe as a self-similar structure, with the same physical
laws appearing at different scales$^{[5]}$.\footnote{Some years ago, it
was shown that strong gravity can be obtained as QCD approximation for
the hadron IR region$^{[6]}$. This makes us argue about the possibility
of extending the scale invariance conjecture to QCD itself.}

Nevertheless, this beautiful picture of nature apparently breaks down
for the quantum behavior of micro-cosmos: the introduction of Planck's
constant defines a very particular scale at which, distinct from large
scales, the quantum effects must be considered. In mathematical
language, we can say that quantum equations, as Schroedinger and Dirac
ones, are not scale invariant, due to the presence of $h$.

The main purpose of this paper is to explore the
picture mentioned above, extending the scale invariance to quantum
behavior. The price to pay, as just discussed, is the scaling of
Planck's constant appearing in quantum equations, leading to the
quantization of large structures, treated till now as classical
ones. This procedure may seem to be rather speculative in character, but
it leads to so impressive coincidences that we need ask ourselves what
truth it contains.

A further purpose is to show that
there seems to exist another, intermediate, scale of invariance besides
that considered by the large numbers hypothesis. As will be shown, the
new scaling leads from particles to typical stars and black holes, in
the same way that the original scaling leads from particles to the
observed universe.

\newpage

\section{The large scale quantization}

${ }$

The cosmological quantities $M$, $R$, and $T$ can be related to $m$,
$r$ and $t$ through the scale relations

\begin{equation}
\label{1}
\frac{T}{t} = \frac{R}{r} = \left( \frac{M}{m} \right)^{\frac{1}{2}} =
\Lambda
\end{equation}

\noindent with $\Lambda \sim 10^{38-41}$. As mass, time, and length are
all we
need for constructing a complete system of units, relation (\ref{1})
completely defines the scale transformation from particles's world to
cosmological one.

With help of (\ref{1}) we can, for instance, scale $h$ in
order to obtain the scale invariance of quantum equations. From a
simple dimensional analysis we have

\begin{equation}
\label{2}
\frac{H}{h} = \Lambda^3
\end{equation}

\noindent leading to a scaled quantum of action given by $H \sim
10^{81}$ J.s, if we choose $\Lambda \sim 10^{38}$.

What is the meaning of this quantization? A possible answer is that
the angular momentum of a rotating universe must be
of the order of $H/2\pi \sim 10^{80}$ J.s.\footnote{Relation (\ref{2})
was already used by Caldirola, Pavisic, and Recami to obtain, from
Planck's constant, the angular momentum of the rotating universe$^{[3]}$.
It is important to note that, choosing $\Lambda \sim 10^{38}$, we are
relating cosmos with typical hadrons$^{[5]}$, what differs from the Dirac
original conjecture (that uses $\Lambda \sim 10^{39}$).}

There is yet no conclusive evidence that universe rotates, although
some speculations about indirect evidences can be made, as the
rotation of galaxies and their clusters$^{[7]}$ and the intergalactic
magnetic field$^{[8,9]}$. But the
important point here is that, {\it if} universe rotates, it should do that
with angular momentum of the order of $H/2\pi$, close to Godel's spin --
the spin value for the rotating cosmological solutions of Einstein's
gravitational equations$^{[7]}$ -- and to the Kerr limit for the spin
of a rotating black-hole with mass of order $10^{50}$ kg. It is also
important to note that this order of magnitude for universe's angular
momentum is within the limits for global rotation obtained from the
cosmic microwave background anisotropy$^{[10]}$ 
and is close to the value derived from the observed rotation of the plane
of polarization of cosmic electromagnetic radiation$^{[11-13]}$.

Besides, (\ref{1}) and (\ref{2}) fit well with Muradian's Regge-like
relation for galaxies and clusters$^{[9]}$, $J = \hbar (M/m)^{3/2}$,
where, here, $M$ stands
for the mass of the object under consideration. It is easy to see that
this relation is in accordance with (\ref{1}) and (\ref{2}) when $M$
is the mass of universe.

\section{The intermediate scale invariance}

${ }$

We can infer another, intermediate, scale of
quantization, related to the angular momenta of stars, which values
concentrate around the order $H'/2\pi \sim 10^{42}$ J.s$^{[9]}$, close to 
the
Kerr limit for a rotating black-hole with mass $M' \sim 10^{30}$ kg.

With these values we calculate the scale factor

\begin{equation}
\label{3}
\lambda \equiv \frac{R'}{r} = \frac{T'}{t} = \frac{H'}{h} \left(
\frac{M'}{m} \right)^{-1} \sim 10^{19}
\end{equation}

\noindent where the first equality comes naturally from the Lorentz
invariance, while the second is obtained, again, from a simple
dimensional analysis.

Besides, with these values for $\lambda$, $M'$ and $H'$, we can infer the
relation

\begin{equation}
\label{4}
\lambda = \left( \frac{M'}{m} \right)^{\frac{1}{3}} = \left(
\frac{H'}{h} \right)^{\frac{1}{4}}
\end{equation}

\noindent which, together (\ref{3}), completely defines the new scale
transformation. Equation (\ref{4}), on the other hand, is in
accordance with Muradian's Regge-like relation for stars and 
planets$^{[9]}$,
$J = \hbar (M/m)^{4/3}$.

From (\ref{3}), we can estimate the values of the $\lambda$-scaled
quantities $R'$ and $T'$ and try to find some physical meaning for
them. From Fermi's scale we obtain $R' \sim 10^4$ m and $T' \sim
10^{-4}$ s. The first can be compared with the gravitational radius of
a typical star: with $M' \sim 10^{30}$ kg it comes $r_g = 2GM'/c^2
\sim 10^3$ m. The second can be compared with the collapse proper-time of
the star, $\tau \sim r_g/c \sim 10^{-5}$ s.

Let us try to understand why the scaled quanta of action is close to
the Kerr limit for the angular momenta of rotating black-holes, in
both ($\Lambda$ and $\lambda$) cases.

Equations (\ref{1})-(\ref{4}) can be put together in the unified form

\begin{equation}
\label{5}
\frac{R_n}{r} = \left( \frac{M_n}{m} \right)^{\frac{1}{n}} = \left(
\frac{H_n}{h} \right)^{\frac{1}{n+1}}
\end{equation}

\noindent with $n=2$ in the $\Lambda$-case and $n=3$ in the
$\lambda$-one. From dimensional analysis, we obtain for the
corresponding gravitational constants

\begin{equation}
\label{6}
G_n = g \left( \frac{M_n}{m} \right)^{\frac{1}{n} - 1} 
\end{equation}

\noindent where $g$ is the strong gravity constant.

Equating the Kerr limit $J_n^{Kerr} = G_n M_n^2/c$ to $H_n/2\pi$ given by
(\ref{5}) and using (\ref{6}), we arrive at the interesting result
$gm^2/\hbar c = 1$. Thus, the coincidence between $H_n/2\pi$ and
$J_n^{Kerr}$ can be based on the fact that the strong structure constant
is of order of unity. Or, reversing the thought, it shows that hadrons
can be considered as maximally rotating black-holes.

It is important to note that the intermediate
scale of length and time is equal to the geometrical average between
Fermi's and cosmological scales. In fact, $(Rr)^{1/2} = r\Lambda^{1/2}
= r \lambda = R'$. It is this fact that guarantees the uniqueness of
the gravitational constant, no matter whether we are dealing with stars or
clusters of galaxies. Indeed, if $G'$ and $G$ are the gravitational
constants at, respectively, $\lambda$- and $\Lambda$-scales, we have,
from (\ref{6}), $G/G' = (G/g)(g/G') = \lambda^2/\Lambda = 1$.

\section{The quantum approach for solar system}

${ }$

Up to now we have only considered orders of magnitude, which, alone,
cannot provide a solid enough basis for the large scale quantization
conjecture. Nevertheless, in a recent paper$^{[14]}$
Oliveira Neto (and, more recently, Agnese and Festa$^{[15]}$) has
presented impressive quantal results concerned to the
solar system, in very good quantitative accordance with the
observational data.\footnote{My thanks to M. Moret for advising me about
Oliveira Neto's work and to a referee for calling my attention to the
paper
by Agnese and Festa.}

For circular orbits, the Newtonian law of gravitation gives $v^2=GM/r$,
where $v$ is the orbital velocity, $M$ is the mass of Sun and $r$ is
the radius of the orbit. Substituting this equation in the
$\lambda$-scaled Bohr quantization condition $L=mvr=nH'/2\pi$ ($m$ is
the mass of the planet), we have

\begin{equation}
\label{7}
r = \frac{n^2H'^2}{4\pi^2GMm^2}
\end{equation}

Agnese and Festa$^{[15]}$ have fitted all the planetary orbits with the
relation $r = n^2 r_1$, with $r_1 = 0.0439$ a.u. So, with $G = 6.67
\times 10^{-11}$ m$^3$/kg.s$^2$, $M = 1.99 \times 10^{30}$
kg and $m = 2.10 \times 10^{26}$ kg
(the average mass of the planets of the solar system),
we find, from (\ref{7}), $H' = (4\pi^2GMm^2r_1)^{1/2} = 1.2 \times
10^{42}$ J.s, that is, the scaled quantum of action obtained, from
(\ref{3}) and (\ref{4}), in the context of the intermediate scale
invariance conjecture.

\section{The redshift quantization}

${ }$

An statistical analysis of astronomical data has
suggested the quantization of cosmic redshifts$^{[16-21]}$, a fact
that has not been explained in the context of the standard
cosmological model. For galaxies, the data has shown a step
of quantization between $cz = 24$ km/s and $cz = 72$ km/s$^{[16,17]}$;
or, from another analysis, between $cz = 6,4 \times 10^3$ km/s and $cz =
1,28 \times 10^4$ km/s$^{[18]}$. These results are also corroborated, at
least on the qualitative level, by the observation that galaxies tend to
cluster in sharp walls, leaving vast
regions devoid of them$^{[19]}$. We shall now try to
establish a possible connection between such observations and the
large scale quantization conjecture, with help of some natural assumptions.

If galaxies are considered as
freely moving in a flat space-time, it is natural to assume
a superior limit for their momenta given by $Mc$, where $M$ is the mass
of universe. This limitation of the space of momenta of galaxies leads,
through the large scale uncertainty relations, to the quantization of their
space-time, with a quantum of length given by $\Delta r \sim H/Mc$. Using
$H \sim 10^{81}$ J.s and $M \sim 10^{50}$ kg, we arrive at
$\Delta r \sim 10^{23}$ m, which corresponds to a velocity step given by
$\Delta v \sim 100$ km/s.

\section{Concluding remarks}

${ }$

Although the curious results shown in this paper, the large scale 
quantization, if genuine, needs a theoretical explanation. We are
probably far away of such a theory, but some remarks can be made in this
direction.

A possible explanation for the quantization of large structures could be
based
on an evolutional point of view: the quantal nature of the universe during
its
initial times (when it had Fermi's scale) has molded its -- apparently
quantized -- nowadays large scale structure. This hypothesis may be
resonable for the case of galaxies and clusters. But it is very improbably
that
intermediate structures like stars and the Solar System maintain the memory 
about the initial conditions. Besides, it would be necessary to explain
the existence
of two different scales of quantization, which does not seem to be very
simple.

Another line of reasoning is to explain the various faces of large
quantization in a fragmented way, in the context of different classical
approachs. As examples, we can mention the ``oscillating universe"
models$^{[22,23]}$, introduced to explain the redshift periodicity of
galaxies. Or Nottale's ``quantum-mechanical" model for solar
system$^{[24]}$, obtained as a diffusion process based on the chaotic
character of the planetary orbits$^{[25]}$. Though distinct from the
approach presented here, Nottale's
model also uses a scaled Planck constant of order $10^{42}$
J.s.\footnote{I thank PR Silva for calling my attention to Nottale's
book.}

Finally, what seems to be the more drastic philosophy: to see the
universe,
including its quantal behavior, as indeed self-similar and to incorporate
this
feature into any fundamental description of the physical world.

\section*{Acknowledgements}

${ }$

I would like to thank H. Arp, R. Muradian, and E. Recami for useful
suggestions. My thanks also to M.C. Nemes for the reading of the manuscript.


\begin{thebibliography}{99}

\bibitem{aaa} Dirac PAM, Nature 139 (1937) 323.

\bibitem{bbb} Salam A and Strathdee J, Phys.Rev.D 16 (1977) 2668; D18
(1978) 4596.

\bibitem{ccc} Caldirola P, Pavisic M and Recami E, Nuov.Cim. 48B
(1978) 205.

\bibitem{ddd} Sivaram C and Sinha KP, Phys.Reports 51 (1979) 111.

\bibitem{eee} Recami E et al, {\it Micro-Universes and Strong
Black-Holes}, in {\it Gravitation: The Space-Time Structure (Proceedings
of Silarg-VIII)}, edited by W.A. Rodrigues et al (World Scientific,
Singapore, 1994), p. 355; and references therein.

\bibitem{fff} Ne'eman Y and Sijacki Dj, Phys.Lett.B 276 (1992) 173.

\bibitem{ggg} Godel K, Rev.Mod.Phys. 21 (1949) 447.

\bibitem{hhh} Surdin M, Phys.Essays 8 (1995) 282.

\bibitem{iii} Muradian R, Astrophys.SpaceSci 69 (1980) 339.

\bibitem{jjj} Kogut A, Hinshaw G and Banday AJ, Phys.Rev.D 55 (1997) 1901.

\bibitem{kkk} Nodland B and Ralston JP, Phys.Rev.Lett. 78(1997)3043.

\bibitem{lll} Obukhov YN, Korotky VA and Hehl FW, astro-ph/9705243.

\bibitem{mmm} Kuhne RW, astro-ph/9708109.

\bibitem{nnn} Oliveira Neto M, Ciência e Cultura 48 (1996) 166.

\bibitem{ooo} Agnese AG and Festa R, Phys.Lett.A 227(1997)165.

\bibitem{ppp} Tifft WG and Cocke WJ, Astroph.J. 287 (1984) 492; 336(1989)
128.

\bibitem{qqq} Guthrie BNG and Napier WM, Mon.Not.R.Astr.Soc. 243 (1990)
431; 253 (1991) 533.

\bibitem{rrr} Broadhurst TJ, Ellis RS, Koo DC and Szalay AS, Nature 343
(1990) 726. 

\bibitem{sss} Geller MJ and Huchra JP, Science 246 (1989) 897.

\bibitem{ttt} Karlsson KG, Astron.Astrophys. 58 (1977) 237.

\bibitem{uuu} Arp H, Bi HG, Chu Y, and Zhu X, Astr.Astrophys. 239 (1990)
33.

\bibitem{vvv} Morikawa M, Astrophys.J.Lett. 362 (1990) L37; Astrophys.J.
369 (1991) 20.

\bibitem{www} Hill CT, Steinhardt PJ and Turner MS, Phys.Lett.B 252
(1990) 343.

\bibitem{xxx} Nottale L, {\it Fractal Space-Time and Microphysics:
Towards a Theory of Scale Relativity} (World Scientific, Singapore, 1993)
pp. 311-321.

\bibitem{yyy} Laskar J, Nature 338 (1989) 237.

\end{thebibliography}
\end{document}